\documentclass[%
  journal=jctcce
]{achemso}

\usepackage[utf8]{inputenc}
\usepackage[T1]{fontenc}
\usepackage{xcolor}
\usepackage[%
  colorlinks=true,
  allcolors=blue
]{hyperref}
\usepackage{url}
\usepackage{enumitem}
\usepackage{amsfonts}
\usepackage{amssymb}
\usepackage{amsmath}
\usepackage{mathtools}
\usepackage[ruled,vlined]{algorithm2e}
\usepackage{lmodern}
\usepackage[normalem]{ulem}
\usepackage{afterpage}

\DeclarePairedDelimiter\ppar{(}{)}              

\newcommand{\dd}[1]{\operatorname{d#1}}

\newcommand{\e}{\operatorname{e}}
\newcommand{\kT}{k_{\mathrm{B}}T}
\newcommand{\kb}{k_{\mathrm{B}}}

\newcommand{\ki}{K_{\mathrm{i}}}

\title{\Large Decoding Binding Pathways of Ligands in Prolyl Oligopeptidase}

\author{Katarzyna Walczewska-Szewc}
\email{kszewc@umk.pl}
\author{Jakub Rydzewski}
\affiliation{%
  Institute of Physics,
  Faculty of Physics, Astronomy and Informatics,
  Nicolaus Copernicus University,
  Grudziadzka 5, 87-100 Toru\'n, Poland
}

\begin{document}

\begin{tocentry}
  \begin{center}
    \includegraphics{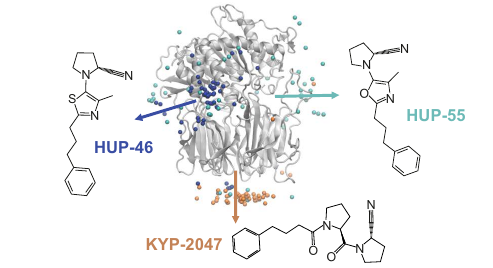}
  \end{center}
\end{tocentry}

\newpage

\begin{abstract}
Neurodegenerative diseases, such as Alzheimer's and Parkinson's, pose a growing global health burden. Prolyl oligopeptidase (PREP) has emerged as a potential therapeutic target in these diseases. Recent studies have shown that direct interaction between PREP and pathological proteins, such as $\alpha$-synuclein and Tau, influences protein aggregation and neuronal function. While most known PREP inhibitors primarily target its enzymatic functions, a new class of ligands, known as HUPs, specifically modulates protein-protein interactions (PPIs), which are crucial in the pathology of neurodegenerative diseases. These structurally distinct ligands exhibit diverse binding behaviors, highlighting the importance of understanding their binding pathways.
In this study, we analyzed the binding pathways and stability of structurally diverse ligands using molecular dynamics simulations and enhanced sampling techniques. Traditional inhibitors, such as KYP-2047, target the active site between the catalytic domains of PREP and the $\beta$-propeller domain, while HUP ligands bind to alternative regions, such as the hinge site, potentially disrupting non-enzymatic PPIs. Using a PLUMED module called \texttt{maze}, we demonstrated that structural variations among ligands lead to distinct binding and unbinding pathways. Free-energy profiles from umbrella sampling revealed key kinetic bottlenecks and differences in pathway selection. For example, HUP-55 exhibits pathway hopping, characterized by diffuse exploration of binding regions before selecting an exit, while KYP-2047 strongly prefers the central tunnel of the $\beta$-propeller domain even under perturbations.
These results suggest that the dynamic interaction between ligands and PREP plays a critical role in their mechanism of action. The ability of HUPs to interact with multiple binding sites and adapt to PREP's conformational changes may be essential for their PPI-targeting effects. This work highlights the need to consider both binding pathways and ligand dynamics in the design of next-generation ligands for PREP and related targets.
\end{abstract}

\maketitle

\newpage

\section{Introduction}
Neurodegenerative diseases, such as Alzheimer's disease (AD) and Parkinson's disease (PD), pose significant challenges to our aging society. With the increasing prevalence of these conditions, there is an urgent need to develop effective therapeutic interventions~\cite{forman2004neurodegenerative,skovronsky2006neurodegenerative}. Prolyl oligopeptidase (PREP), an enzyme implicated in various neurodegenerative disorders, has emerged as a promising target for this reason~\cite{myohanen2012prolyl,vanderveken2012p2}. PREP, initially identified as a serine protease involved in neuropeptide metabolism, has been associated with neurodegenerative diseases such as AD, PD, and others. Previous research has shown that the abnormal activity of PREP is associated with cognitive decline and dementia in animal models, making it an attractive target for therapy~\cite{brandt2007suggested,williams2002common}. Currently, the primary focus is on non-enzymatic protein-protein interactions (PPI) related to PREP functions. Recent studies have revealed that direct interaction between PREP, $\alpha$-Synuclein ($\alpha$Syn), and Tau results in increased aggregation of such proteins~\cite{brandt2005search,brandt2008prolyl,savolainen2015prolyl,svarcbahs2019new,etelainen2023prolyl,vanelzen2025prolyl}. Anomalous processing and aggregation of $\alpha$Syn and Tau are considered the main factors in cellular toxicity in AD and PD. Furthermore, PREP interactions with protein phosphatase 2A (PP2A) decrease PP2A activity and autophagy, further advancing these neurodegenerative diseases~\cite{etelainen2021prolyl}.

Most known PREP ligands have been designed to inhibit its enzymatic functions, targeting the well-characterized active site located at the interface of the two primary domains of PREP -- an $\alpha$/$\beta$-catalytic domain and a $\beta$-propeller domain composed of seven $\beta$-sheet blades. This active site contains a catalytic triad of S554, H680, and D641. In some cases, inhibitors also modulate PPI-related effects, although this is often an unintended side effect rather than the primary goal~\cite{kilpelainen2020effect}. Recently developed HUP ligands have been optimized to selectively regulate the nonenzymatic functions of PREP~\cite{kilpelainen2023nonpeptidic}. Such ligands bind deeper within the hinge region, connecting the two main domains of PREP~\cite{patsi2024amino}. However, their broader impact on the protein, particularly the mechanisms by which they modulate PPI properties, remains unclear.
\begin{figure}
    \centering
    \includegraphics[width=0.9\linewidth]{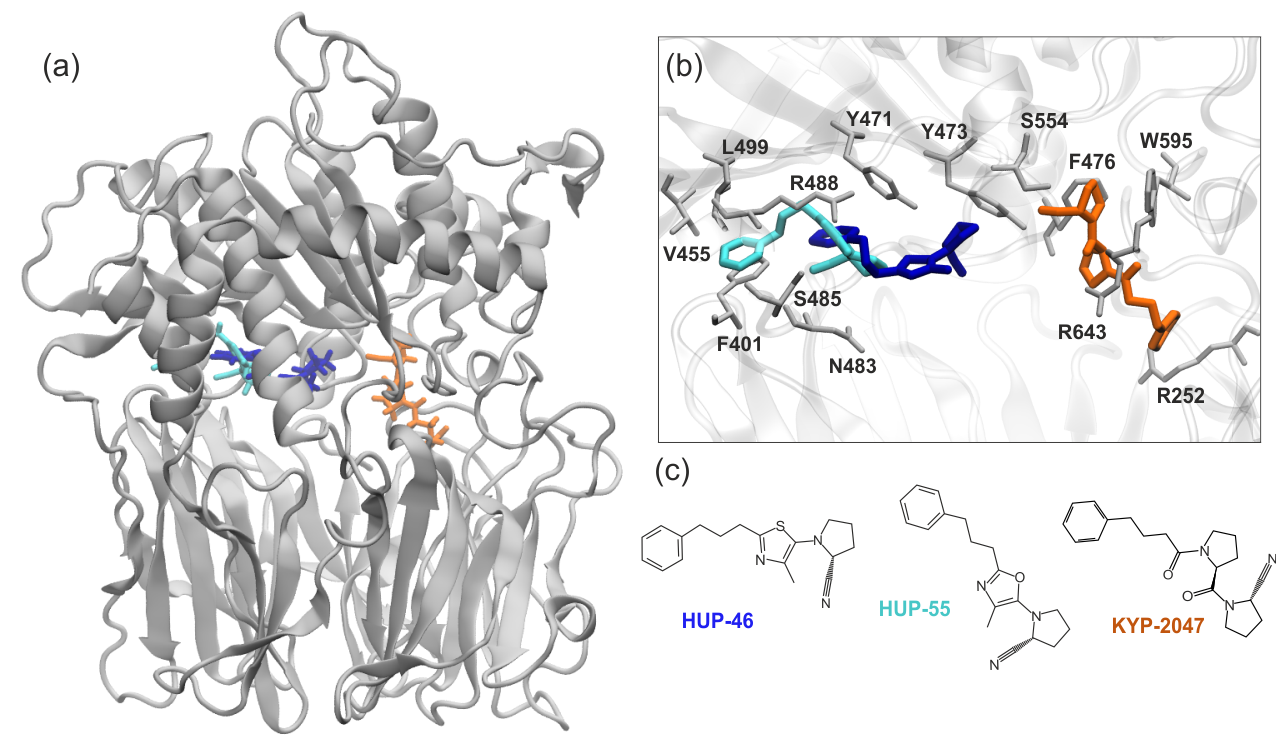}
    \caption{(a) PREP structure with the investigated ligands: HUP-46 (blue), HUP-55 (cyan), and KYP-2047 (orange). (b) Main interactions between PREP's amino acids and the ligands in their binding sites. (c) Chemical structures of the compounds.}
    \label{fig:intro}
\end{figure}

The emergence of new drugs targeting alternative binding sites, in general, raises questions about their ligand-binding properties~\cite{carroll2012evidence,baron2013molecular,rydzewski2017ligand,ribeiro2018kinetics,bernetti2019kinetics}, particularly regarding diverse and potentially multiple binding pathways, which have recently been shown to influence rate-limiting processes during ligand-protein dissociation~\cite{tiwary2015kinetics,rydzewski2016machine,rydzewski2017thermodynamics,rydzewski2018kinetics,rydzewski2019finding,lotz2018unbiased}. The differences in the effectiveness of these two structurally distinct groups of ligands may arise from their interaction sites with PREP and how they associate with the protein along multiple binding pathways~\cite{walczewska2024structural}. A recent hypothesis suggests that known ligands enter the cavity between the two subunits via flexible loops referred to as loop A (residues 189–209), loop B (residues 577–608), loop C (residues 636–646), and the His loop (residues 676–685)~\cite{tsirigotaki2017dynamics,szeltner2004concerted}. The access to the interior of the protein depends on the local conformations of its loops, which may be open or closed at any given time. An earlier idea proposed that ligands and substrates might enter through the $\beta$-propeller pore~\cite{fulop1998prolyl,fulop2000catalysis}. Understanding how ligands bind and unbind from the protein could be key to achieving favorable inhibitory/regulatory characteristics, such as low IC50 values~\cite{tummino2008residence}.

Previous studies on PREP ligand binding pathways have focused primarily on peptide-mimicking inhibitors that target the active site. Multiple binding pathways have been explored in one such study, with particular emphasis on the $\beta$-propeller domain and the flexible loop interface among loops A, B, C, and His~\cite{stpierre2011umbrella}. This suggests that the flexible loop pathway is the most probable route, although it would require large-scale domain reorientation~\cite{kaszuba2012molecular,stpierre2011umbrella,szeltner2013loops,tsirigotaki2017dynamics}. However, the transition of human PREP from the closed to open conformation appears to occur frequently enough to allow the ligand access to the protein~\cite{ellis2019crystal,tsirigotaki2017dynamics,walczewska2022inhibition}. Another approach has also identified two possible ligand entry routes: one via the $\beta$-propeller pore and another through the loop region~\cite{kotev2015unveiling}.

To gain insight into the ligand-binding pathways in PREP, we comprehensively compared various types of compounds. First, we identified their binding sites and assessed their stability within these regions. Next, we demonstrated that slight differences in the structure of ligands and their chemical properties lead to completely different binding pathways in PREP. Lastly, the binding pathways were examined using umbrella sampling to quantify the corresponding free-energy profiles. This approach enabled us to identify the key kinetically limiting steps involved in these binding transitions.

\section{Methods}
\subsection{Models}
The structure of human PREP~\cite{haffner2008pyrrolidinyl} (PDB ID: 3DDU) was processed using Maestro within Schrodinger~\cite{madhavi2013protein}, employing the OPLS3e force field~\cite{roos2019opls3e} and PROPKA at pH 7.4. We also prepared four mutated systems to investigate the potential effect of mutations on the identified binding pathways and assess the stability of the complexes. The structures of three ligands, KYP-2047, HUP-46, and HUP-55, underwent geometry optimization through quantum chemical calculations using the ORCA 5.0.3 program~\cite{neese2020orca} using DFT-D3. The resulting ligand structures were used as input to generate CHARMM36m force field parameters~\cite{huang2017charmm36m} via the Swissparam server~\cite{zoete2011swissparam,bugnon2023swissparam}. Flexible docking of the ligands was performed using Glide Schrodinger~\cite{friesner2006extra}. After selecting the optimal conformation, each protein-ligand complex was immersed in a water box with dimensions of 10 nm in each direction and neutralized by adding Na$^+$ and Cl$^-$ ions (0.15 M concentration). These systems comprised approximately 100,000 atoms. This protocol resulted in the following seven complexes: PREP with the KYP-2047, HUP-46, and HUP-55 ligands; mutated PREPs with HUP-46 (L94 and I690C, T68C and T686C); and mutated PREPs with KYP-2047 (Q397C,  Q397C and C255T).

\subsection{Molecular Dynamics Simulations}
All MD simulations were run using the Gromacs 2021.3 software~\cite{gromacs} patched with PLUMED 2.8~\cite{plumed,plumed-nest,plumed-tutorials}. Each protein-ligand complex was initially fixed while water and ions underwent equilibration for 0.5 ns. Following this, three MD simulations of 0.5 ns each were conducted in the NVT ensemble at 310 K, applying position restraints of varying magnitudes (1000, 500, and 100 kJ/(mol$\cdot$ nm$^2$)) to the protein C$\alpha$ atoms and the ligand. Subsequently, a fourth MD simulation with a duration of 1 ns was performed in the NPT ensemble, applying a harmonic restraint of 10 kJ/(mol$\cdot$ nm$^2$) to the protein C$\alpha$ atoms and ligand while maintaining a constant pressure of 1 atm using the Berendsen barostat. Finally, an unrestrained equilibration MD simulation lasting 50 ns was conducted in the NPT ensemble, with constant pressure of 1 atm maintained using the Parrinello–Rahman barostat. The further production runs were simulated through 500 ns with the same parameters as the last step of equilibration. The temperature was held constant at 310 K by applying the velocity rescaling thermostat~\cite{vrescale}. A time step of 2 fs was used. Bonds involving hydrogen atoms were constrained using LINCS~\cite{hess2008p}. We refer to Supplementary Information for a detailed list of all MD simulations performed in this study (Tab. S1). 

To identify binding pathways, we used an adaptive biasing method implemented in the \texttt{maze} module~\cite{rydzewski2019finding,rydzewski2020maze} of PLUMED~\cite{plumed,plumed-nest,plumed-tutorials}. Following our previous protocol~\cite{plumed-tutorials}, for each complex (HUP-46-, HUP-55-, KYP-2047-PREP, and HUP-46-KYP-2047- and the mutated PREPs), we ran 50 MD simulations in the NVT ensemble at 310 K. The ligands were pulled with a constant velocity of 0.00035 nm/ps from PREP by minimizing a loss function that described the contacts between the inhibitors and proteins. The minimization was launched every 10 ns using simulated annealing. The loss function was $Q = \sum_{kl} \exp(-r_{kl})/r_{kl}$, where $r_{kl}$ is the distance between the $k$-th atom of PREP and the $l$-th atom of the ligand, with a cutoff for $r_{kl}$ set to 0.7 nm. The simulations were terminated when the ligands dissociated from PREP or when the simulation exceeded 200 ns.

We selected representative binding pathways and used them to define collective variables (CVs) for umbrella sampling simulations~\cite{torrie1977nonphysical,Kastner2011umbreallsampling}. The pathways were clustered based on their exit points using the DBSCAN algorithm. For each cluster, we identified the most representative pathway as the one closest to the cluster centroid (see Fig. S4 in Supporting Information). These representatives were then visually inspected to ensure they were reasonable compared to other pathways, and a subset of pathways was chosen based on this evaluation. Each binding pathway was divided into frames, with the center of ligand mass traveling approximately 0.2 nm between consecutive frames. Depending on the length of the pathway, this resulted in 18 to 24 frames per CV, defined as the center-of-mass distance between the bound ligand and the dissociating ligand. We ran a separate 10-ns MD simulation for each window, restraining the ligand with a harmonic potential of 500 kJ/(mol$\cdot$ nm$^2$). The free-energy profiles were calculated using the weighted histogram analysis method~\cite{hub2010g_wham} implemented in Gromacs. Error bars were determined with bootstrapping. For each case, the free-energy difference between the bound (B) and unbound (U) states was estimated as $\Delta F = -\frac{1}{\beta} \log\ppar{{{\int_{\mathrm{B}}\dd{z}\e^{-\beta F(z)}}/{\int_{\mathrm{U}}\dd{z}\e^{-\beta F(z)}}}}$, where $\beta=1/\kT$ is the inverse temperature with temperature $T$ and Boltzmann's constant $\kb$, and $z$ denotes a CV.

We used the MDAnalysis Python library for analysis~\cite{mdanalysis2011}. Molecular visualizations were created using VMD~\cite{humphrey1996vmd}.

All the data and PLUMED input files required to reproduce the results reported in this paper are available on PLUMED-NEST, the public repository of the PLUMED consortium~\cite{plumed-nest}, as plumID:XX.YYY and on RepOD repository (\url{https://doi.org/10.18150/ZUISNK}). The implementation of the \texttt{maze} module for PLUMED is available in a git repository (\url{https://github.com/jakryd/plumed2-maze}).

\section{Results}
\subsection{Binding Sites}
Our results demonstrate different binding site preferences among the selected ligands. The classical peptide-like inhibitor KYP-2047, which affects catalytic functions and PPIs, binds predominantly to the well-known and characterized binding site. Previous studies have shown that KYP-2047 can form a covalent bond with S554 in the active center of the enzyme~\cite{kaszuba2012molecular} (see Fig. \ref{fig:intro}b). Although this covalent bond was not included in our model, the ligand maintains significant stability, occupying the same binding site as its covalently bound counterpart~\cite{walczewska2022inhibition,kaszuba2012molecular}. The positioning of KYP-2047 within the binding cavity is regulated by interactions involving two carbonyl groups with R643 and W595, demonstrating a near 100\% contact frequency throughout the eight sets of 0.5 $\mu$s simulations (see Tab. S1 in Supporting Information). The part of the ligand containing the CN functional group exhibits frequent interactions with residues S554, F476, W595, and Y473. Furthermore, the octahydroindole moiety of the ligand frequently interacts with C255, F476, W595, and R643. The aromatic benzene ring engages in a series of hydrophobic interactions with F173, M235, and I591, further enhancing the binding of the ligand.

Two additional ligands, belonging to the newly developed HUP group, have been confirmed by recent studies~\cite{patsi2024amino} through MD simulations and mutagenesis to bind at an alternative site located approximately 1.8 nm away from the active site toward the hinge region. This alternative binding site of PREP appears to be more significant than the primary active site for modulating its PPI-mediated functions. For the 5-aminooxazole-based HUP-55, after docking to a specific binding site similar to that reported by Pätsi et al. (see Fig.~\ref{fig:intro}b and Fig.~S1 in Supporting Information), the ligand remained stable for nearly 500 ns of MD simulation. The stability of HUP-55 binding is mainly attributed to interactions between the oxazole ring and residues Y471 and R488. Additionally, our results suggest that the CN functional group engages with S485 and N483, while hydrophobic interactions with V455, L499, and F401 help stabilize the benzene ring. However, in three out of five simulations, the ligand diffuses out of the binding site around 500 ns, suggesting that this binding is not as stable as KYP-2047 in the active site. 

The 5-aminothiazole-based HUP-46, despite its clear preference for the alternative binding site, resides there for approximately 200 ns before diffusing into the internal cavity of PREP (see Fig. S1 in Supporting Information). Interestingly, the ligand frequently returns to the alternative binding site, suggesting a dynamic and transient interaction rather than stable binding.

These observations highlight a potential mechanism of action for HUP ligands, characterized by greater flexibility in the selection of binding sites compared to standard inhibitors like KYP-2047. However, it is unclear whether this is relevant to their role in modulating PPIs. The ability to interact dynamically with multiple binding sites may be due to the adaptation to conformational changes in PREP, which could be critical for their PPI-targeting mechanism.

\subsection{Unbinding Reaction Pathways}
\begin{figure}
    \centering
    \includegraphics[width=0.9\linewidth]{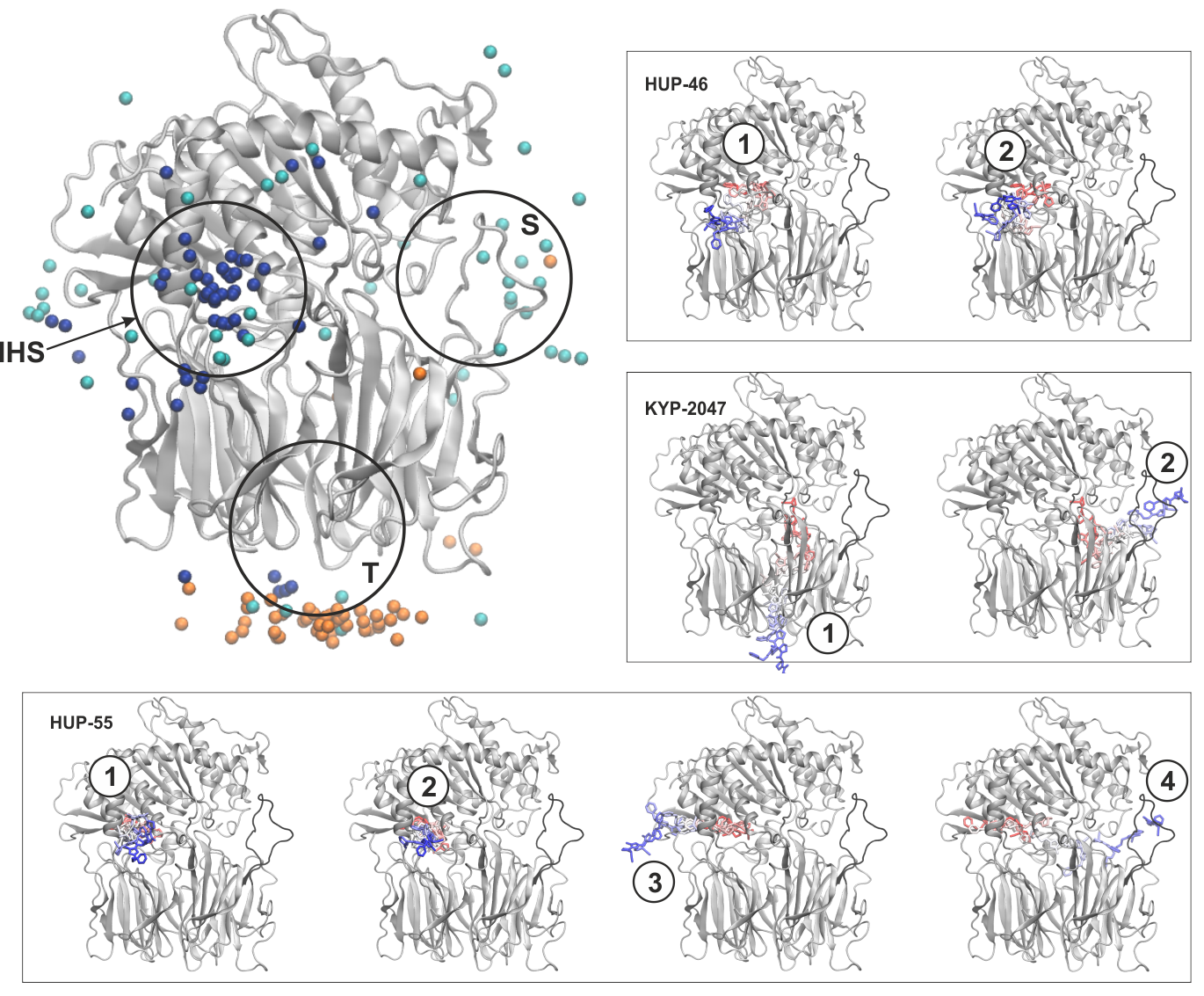}
    \caption{Most frequently chosen exit pathways of the ligand from PREP. Dark blue, cyan, and orange spheres indicate the center-of-mass position of the ligand after leaving the protein, corresponding to HUP46, HUP55, and KYP2047, respectively. Insets for each ligand show specific exit trajectories captured from simulations, representing the most frequently chosen pathways. The ligand is colored red at the beginning and blue at the end of its journey through the protein.}
    \label{fig:exits}
\end{figure}

Starting from the previously identified binding site (the active site in the case of KYP-2047 and alternative site for HUP-46 and HUP-55, see Fig.~\ref{fig:intro}b), we conducted independent ligand unbinding simulations using the \texttt{maze} module implemented in PLUMED. Fig.~\ref{fig:exits}a shows the final positions of the ligands after leaving the central pocket (orange spheres represent KYP-2047, cyan HUP-55, and dark blue HUP-46). Interestingly, we found that the three ligands exhibit distinct exit patterns. The HUP-55 ligand jumps between various possible pathways, exiting through different protein regions as it searches for transient tunnels. In contrast, HUP-46 consistently follows an exit pathway between helices 58:71, 685:705, and Blade1 (residues 92:97), an area we designated ``inter-helical site'' (marked as IHS in Fig.~\ref{fig:exits}). Although HUP-55 occasionally populates this route, its dissociation events are more uniformly distributed as it explores alternative options when the primary binding pathway is temporarily closed. These results suggest that HUP-55 can show a behavior known as pathway hopping, where ligands migrate more diffusively in proteins due to large fluctuations in binding regions before selecting an exit~\cite{lotz2018unbiased}. As recently shown, pathway jumping occurs in the acetylcholinesterase-Huperzine A complex, which is also considered important for neurological disorders~\cite{rydzewski2018kinetics}.

\begin{table}[t]
\footnotesize
\caption{List of the identified unbinding reaction pathways for the PREP-ligand complexes.}
\begin{tabular}{l l l l} 
 \hline
 Protein & Ligand & Pathway & Description \\
 \hline
 PREP & KYP-2047 & 1 & Tunnel in the $\beta$-propeller domain \\ 
      &         & 2 & Side opening through flexible loops: A, B and His \\\hline
 PREP & HUP-46   & 1 & Inter-helical site, the benzene ring toward solvent \\ 
      &         & 2 & Inter-helical site, the pyrrolidine ring toward solvent \\\hline
 PREP & HUP-55   & 1 & Inter-helical site, the pyrrolidine ring toward solvent \\ 
      &         & 2 & inter-helical site,  benzene ring toward solvent \\ 
      &         & 3 & Hinge region \\ 
      &         & 4 & Side opening, through flexible loops: A, B and His \\ \hline
\end{tabular}
\end{table}

The dissociation of HUP-46 through the inter-helical site occurs more quickly, primarily within the first 20 ns of simulation (see Fig. S2 in Supporting Information for the time distribution of dissociation events). In comparison, HUP-55 exits between 20 to 40 ns, although there can be occasional delays of up to 200 ns. The fastest exits for HUP-55 tend to occur near the inter-helical site. Consequently, attempts to use alternative pathways often prolong the process, as the ligand must wait for conformational changes to open these routes, such as the side opening located next to loops A, B, and the His-loop (marked as S in Fig.~\ref{fig:exits}).

Interestingly, KYP-2047 prefers a different exit route than suggested in earlier findings~\cite{tsirigotaki2017dynamics,stpierre2011umbrella}. The average exit times for KYP-2047 are relatively short, generally occurring between 20 and 40 ns (see Fig. S2 in Supporting Information). However, the inhibitor predominantly exits via a tunnel in the $\beta$-propeller domain (marked as T in Fig.~\ref{fig:exits}). This exit route has been previously considered less favorable based on free-energy calculations by St-Pierre et al.~\cite{stpierre2011umbrella}, who have concluded that the corresponding tunnel is too narrow and has insufficient fluctuations to enable dissociation for such inhibitors effectively. However, Kotev et al.~\cite{kotev2015unveiling} have suggested that dissociation through this pathway is possible and can be biologically relevant. The alternative, more challenging exit through the side opening next to loops A, B, and the His-loop generally takes longer, often requiring over 100 ns.

The representative exit pathways for each ligand are shown as colored trajectories in the inset of Fig.~\ref{fig:exits}, with positions shaded from red (trajectory start) to blue (trajectory end). For HUP-46, two exits occur through the inter-helical site with slight variations in the ligand's initial orientation. KYP-2047 shows one exit through the tunnel in the $\beta$-propeller domain and another via the flexible loop region. HUP-55 exhibits two exits through the inter-helical site (varying in ligand's orientation), another through a temporary tunnel near the hinge region, and one through the loop region (A, B, and His-loop).

\subsection{Blocking the Preferred Exit Route of HUP-46}
To further investigate the role of specific exit pathways, we performed additional simulations of ligand unbinding, with a focus on two specific cases. We introduced two sets of mutations in PREP to block the inter-helical exit pathway preferred by HUP-46. In the first system, L94 and I690 were mutated to cysteines, while in the second, cysteines were introduced at T68 and T686 to form a disulfide bridge obstructing the transient tunnel. 

\begin{figure}
    \centering
    \includegraphics[width=0.9\linewidth]{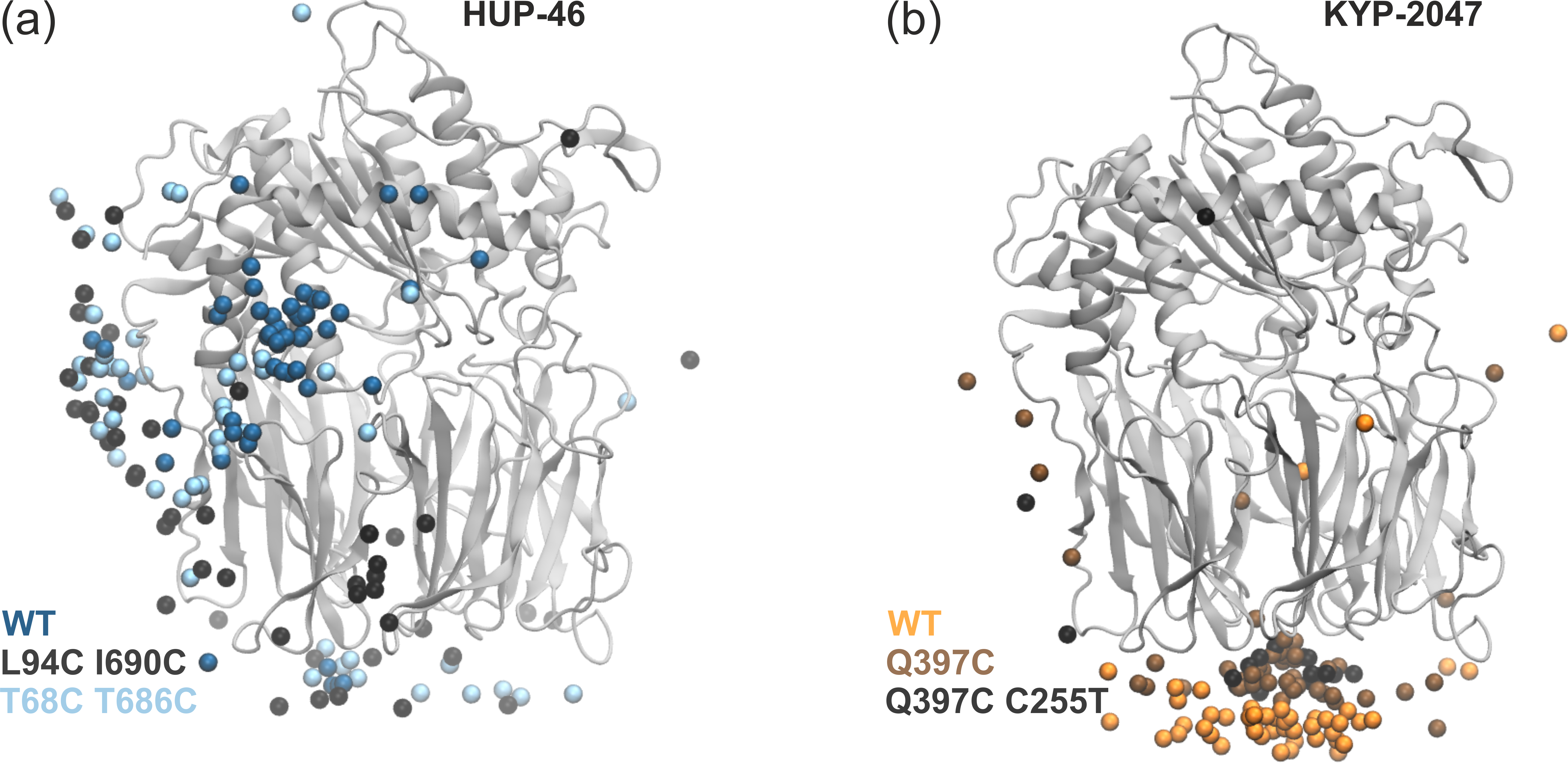}
    \caption{Most frequently chosen exit pathways of the ligand from mutated PREP systems. (a) Dark blue, black, and cyan spheres represent the center-of-mass positions of HUP-46 after exiting the protein in the wild-type PREP (dark blue), L94C/I690C mutated PREP (black), and T68C/T686C mutated PREP (cyan). (b) Orange, brown, and black spheres represent the center-of-mass positions of KYP-2047 after exiting the protein in the wild-type PREP (orange), Q397C mutated PREP (brown), and Q397C/C255T mutated PREP (black).
}
    \label{fig:mut}
\end{figure}

The dynamics of the rest of the protein was minimally affected (see Fig. S3 in Supporting Information). Our simulations of the mutated systems reveal that the preferred exit route through the inter-helical site is no longer accessible, which forces the ligand to seek alternative pathways. In Fig.~\ref{fig:mut}a, the locations of exits for the mutated systems (light blue and black spheres) are more evenly spread than the wild-type PREP (dark blue spheres). Additionally, the time needed to find an exit increased, with average exit times for the mutated systems comparable to those observed for HUP-55 and KYP-2047 (20-40 ns). These findings indicate that such mutations may alter ligand behavior, which could be further verified experimentally.

\subsection{Preference of KYP-2047 for the $\beta$-Propeller Exit Tunnel}
Given the structural arrangement of the $\beta$-propeller domain, where the tunnel is formed by seven ``blades'', we introduced the Q397C mutation to form a disulfide bridge with the native C78. This mutation reduces the flexibility of the blades, limiting their ability to undergo ``breathing'' fluctuations that could otherwise widen the tunnel. To prevent unintended interactions with the native C255, this residue was mutated to threonine in an additional mutated system.

Despite these mutations, KYP-2047 continues to exhibit a strong preference for exiting through the tunnel in the $\beta$-propeller domain (see Fig.~\ref{fig:mut}b, with brown and black spheres representing mutated systems, and orange spheres for the wild-type PREP). Neither the exit preference nor the average exit times are significantly altered by these mutations. These results suggest that the ligand not only fits well through the tunnel but also retains sufficient flexibility to overcome additional steric hindrance introduced by the mutations.

\subsection{Diffusion in Binding Sites for Nonpeptidic HUP Ligands}
To analyze possible reaction pathways, we calculated the free energy required for ligand dissociation along the representative trajectories in each complex. To achieve this, we used umbrella sampling, a method successfully used in previous studies to investigate ligand dissociation from PREP~\cite{stpierre2011umbrella}. 

\begin{figure}
    \centering
    \includegraphics[width=1\linewidth]{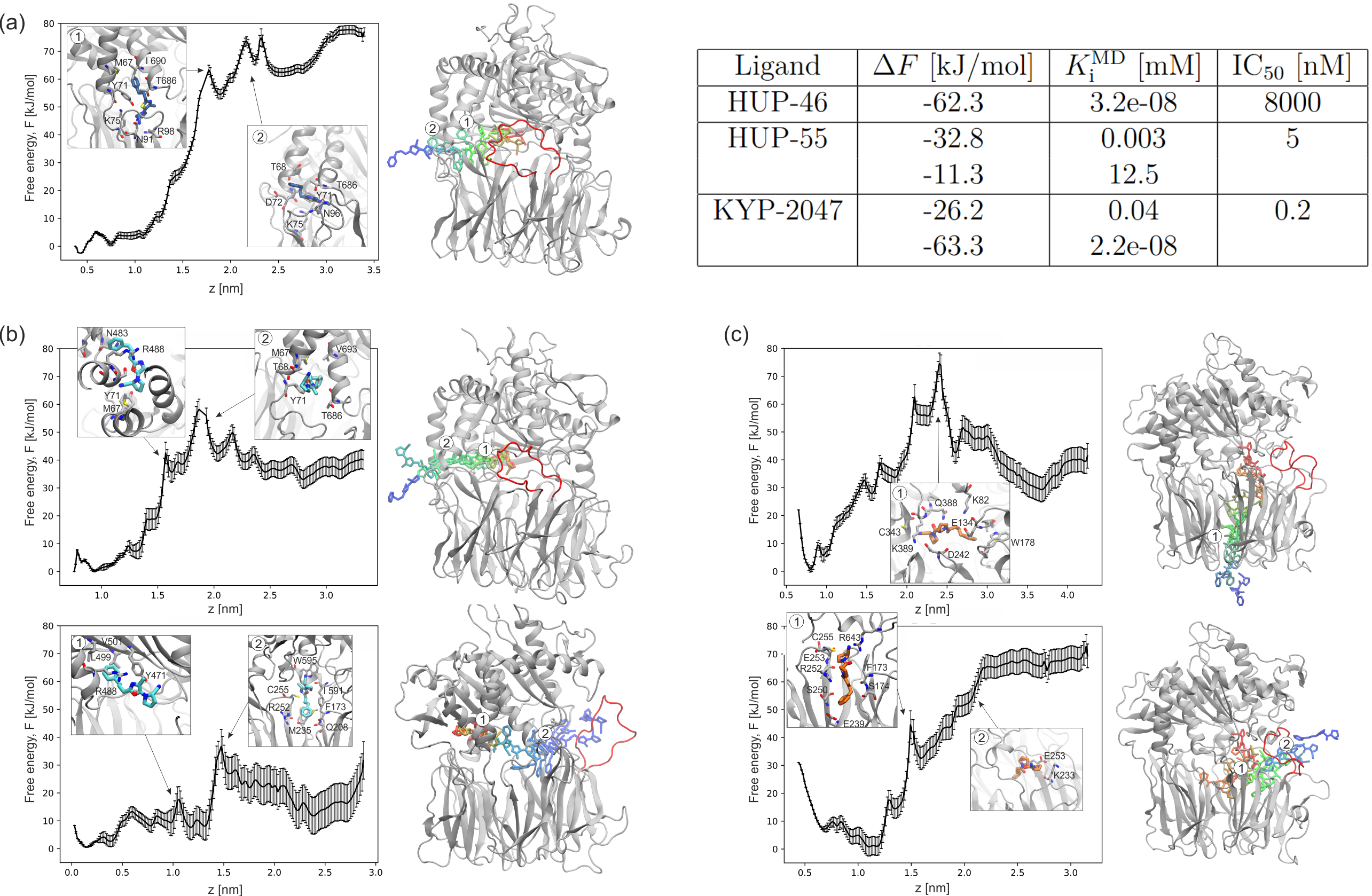}
    \caption{Free-energy profiles depicting ligand unbinding along specific pathways for (a) HUP46, (b) HUP55, and (c) KYP2047. On the right, ligand trajectories are shown, with red structures representing the starting positions and blue structures the final ones. Insets within the free-energy plots highlight key regions where the ligand encounters significant energy barriers. The table summarizes the estimated free-energy difference between the bound and unbound states, $\Delta F$ , the inhibition constant derived from MD simulations, $\ki^{\mathrm{MD}}$, and the experimentally determined IC$_{50}$ values for each ligand. }
    \label{fig:umbrella}
\end{figure}
%

For HUP-46, which favors the exit through the inter-helical site, the free-energy difference between the bound and unbound states is around -62 kJ/mol (Fig.~\ref{fig:umbrella}a). The main kinetic bottleneck during dissociation occurs after the ligand exits its binding site, specifically when detaching from the well-adapted pocket formed along the inner wall of the protein cavity. Here, the benzene ring of HUP-46 resides in a hydrophobic pocket formed by M67, I690, and Y71, while the nitrile group engages with polar and positively charged residues K75, N91, R98, and T686. Upon release of the benzene ring, the ligand is also restrained by interactions with T68 and D72.

For HUP-55, we investigated two possible dissociation pathways in more detail. The first pathway we consider is through the inter-helical site (Fig.~\ref{fig:umbrella}b, top), similar to where HUP-46 exits PREP. We can see that the free-energy difference of around -33 kJ/mol is primarily caused by leaving the binding site of HUP-55 and traversing the sequence of residues that form the ``wall'' of the protein. Notably, the free-energy difference for HUP-55 is lower than that of HUP-46. This raises questions about whether this change is attributed to the structural differences between aminooxazole and aminothiazole or whether it results from fluctuations in the side chains of PREP during the simulation.

The second dissociation pathway for HUP-55 is longer, passing through the internal cavity of the protein to the side exit near the flexible loops (Fig.~\ref{fig:umbrella}b, bottom). This pathway is particularly interesting as it demonstrates that the most challenging steps are associated with leaving two previously characterized binding sites within PREP. The first site, located deep in the hinge region, is typical for HUP ligands, while the second is closer to the active center of PREP. Transitioning from the first binding site to the second proves relatively easy, with an energy difference of just around -11 kJ/mol. This suggests that movement between these two sites is somewhat straightforward and may occur frequently, possibly as part of the natural processing of physiological substrates in PREP. Exiting from the second binding site through the flexible loop region required overcoming an additional energy barrier of approximately 30 kJ/mol.

%
Interestingly, while the pathway for HUP-46 is dominant in the \texttt{maze} simulations, the umbrella sampling calculations revealed relatively high free-energy barriers along the same path. This discrepancy may arise from the inherent differences between the two methods. \texttt{maze} primarily explores sterically accessible routes that might not always align with the most thermodynamically favorable pathways. Thus, \texttt{maze} simulations may overrepresent paths that are structurally easier to traverse but energetically less favorable. 

Another plausible explanation is that other potential pathways were temporarily inaccessible because of protein conformational fluctuations during the simulations. Dynamic elements, such as loops or domains, could sporadically obstruct access to alternative, potentially more favorable routes, like the side opening. This transient occlusion might explain why the inter-helical pathway appeared more frequently in \texttt{maze}, even though it exhibits higher free-energy barriers in umbrella sampling.

Furthermore, the unique interactions of HUP-46 with the protein may favor this specific pathway. The aminooxazole group of the ligand, for example, can form stabilizing hydrogen bonds or electrostatic interactions that restrict its mobility in other directions. This could contrast with ligands like HUP-55, which feature an aminothiazole group, potentially leading to different binding dynamics and preferred pathways.

\subsection{KYP-2047 Unbinding}
The final two dissociation pathways analyzed involve the peptidic ligand KYP-2047, which binds near the active center of PREP (Fig.~\ref{fig:umbrella}c). One of the most commonly chosen and time-efficient exit routes for KYP-2047 is through the tunnel in the $\beta$-propeller domain (Fig.~\ref{fig:umbrella}c, top). The most challenging step along this route is passing through the cluster of charged residues at the entrance of the tunnel (inset 1, the peak around $z=2.5$), which creates an energy barrier of approximately 80 kJ/mol (measured from the baseline). However, the estimated free-energy difference between the bound and unbound states is about -26 kJ/mol, which is relatively small and suggests that, despite the high transition barrier, this pathway may still be the preferred dissociation route.

In the study by St-Pierre et al. on ZPP dissociation pathways in PREP~\cite{stpierre2011umbrella}, this entry route was associated with a transition barrier of approximately -101 kJ/mol, leading to estimates of inhibition constants that were several orders of magnitude lower than the values observed experimentally. As a result, this pathway was deemed nonphysical. However, it is important to note that we have two different ligands and two distinct approaches to select trajectory points for umbrella sampling (US): in our case, \texttt{maze} and, in theirs, SMD. Comparison of ZPP with KYP-2047 requires caution, as the P3 part of the inhibitor differs significantly. Specifically, ZPP features a carbamate group instead of an amide bond, which could be considered to be more bulky and less flexible than the corresponding moiety in KYP-2047.

The second pathway for KYP-2047 involves exiting through the flexible loop region (Fig.~\ref{fig:umbrella}c, bottom). All observed exit trajectories for this pathway require a slight downward shift toward the \textit{$\beta$}-propeller domain. After overcoming the interactions that anchor the ligand just below the binding site (inset 1), the ligand moves through the flexible loop region, utilizing the flexibility of the loops between blades 3 and 4 of the \textit{$\beta$}-propeller domain (inset 2). The free-energy difference for exiting along this pathway is approximately -63 kJ/mol.

In the case of StPierre's work with ZPP~\cite{stpierre2011umbrella}, the free-energy difference between the bound and unbound states was much smaller, around -18 kJ/mol, with a transition barrier of approximately 25 kJ/mol. In our study, both the barrier and the energy differences are significantly higher, primarily because of the challenge of 'pushing' the ligand through the unfavorably positioned loops, which likely hinders dissociation. 

A notable difference can be observed in the initial profiles of the free-energy curves near the starting positions of the ligands. For HUP ligands, the curve is relatively flat, while for KYP-2047, it forms a distinct deep minimum. This could explain the differences in ligand behavior observed during the dynamics of the system: KYP-2047 binds strongly and stably, whereas HUPs exhibit a more dynamic and transient binding mode, reflecting their ability to adapt to conformational changes in PREP and interact with multiple binding sites.

\subsection{Inhibition Constants}
We aimed to compare our results with available experimental data on ligand binding affinities. For KYP-2047, inhibition constant ($\ki$) data is available, quantifying the inhibitor concentration required to occupy half of the binding sites in the enzyme or receptor. Using the formula $\ki = [1\mathrm{M}] \e^{-\beta\Delta F}$, it is possible to estimate $\ki$ based on the free-energy difference between the bound and unbound states~\cite{stpierre2011umbrella}. Applying this to the free-energy differences obtained from our calculated pathways, we estimate $\ki$ to range from 0.02 nM for pathway 2 (side exit) to approximately 0.4 mM for pathway 1 (exit via the $\beta$-propeller domain). The experimental value for KYP-2047 is 0.02 nM, suggesting that our simulations reasonably capture the general binding affinity range. 

For HUP-46 and HUP-55, experimental $\ki$ values are not available; only IC50 values have been reported (8 $\mu$M and 5 nM, respectively), along with EC50 values of 100 nM for HUP-46 and 275 nM for HUP-55~\cite{patsi2024amino}. This presents a significant challenge in interpretation, as IC50 values, which represent the concentration required to inhibit a specific biological activity (typically enzymatic function), are highly dependent on the available enzyme concentration and experimental conditions. Consequently, while HUP-55 can be classified as an inhibitor due to its low IC50 value (5 nM), HUP-46, with an IC50 of 8 $\mu$M, is more likely to function as a ligand rather than a potent inhibitor. However, EC50 values, which quantify the concentration needed to induce a specific biological effect (in this case, modulating the function derived from PPI), may provide a more relevant measure for these compounds, as both HUP-46 and HUP-55 were specifically designed to modulate PPIs.

The $\ki$ values derived from MD simulations for HUP-46 (0.03 nM) and HUP-55 (ranging from 0.003 nM to 12.5 mM) suggest that HUP-46 binds to PREP with nanomolar affinity, placing it among the most potent PREP inhibitors, despite its relatively high IC50 value. In contrast, HUP-55 exhibits a much weaker binding affinity, with inhibition constants spanning the micromolar to the millimolar range. These results can be placed within the range of literature-reported values for other PREP ligands, which typically range from the sub-nanomolar (e.g., KYP-2047 with 0.02 nM) to the micromolar ranges for weaker inhibitors. The nanomolar affinity of HUP-46 suggests strong binding, while the millimolar affinity of HUP-55 indicates significantly weaker binding compared to most known inhibitors. However, it is important to note that these values are highly approximate, as they are derived from specific exit pathways modeled in the simulations. Further experimental validation is required to confirm these predictions.

\section{Conclusions}
This work investigates inhibitor binding and unbinding pathways in prolyl oligopeptidase (PREP), demonstrating how slight structural differences between ligands can lead to distinct binding behaviors and exit routes. We have examined these pathways by combining extensive MD simulations and enhanced sampling techniques and quantified the corresponding free-energy profiles.

The analysis revealed that HUP-55 diffuses out of the binding site after approximately 500 ns, suggesting that its binding is less stable than KYP-2047. In contrast, HUP-46 shows greater flexibility in selecting binding sites, which is unusual for standard inhibitors such as KYP-2047. In particular, the initial profiles of the free-energy curves also reflect these differences: for the HUP ligands, the curve is relatively flat near the starting position, while for KYP-2047, it forms a distinct deep minimum. This observation further supports the idea that KYP-2047 binds strongly and stably, whereas HUPs exhibit a more dynamic and transient binding mode, likely due to their ability to adapt to conformational changes in PREP and interact with multiple binding sites.

The unbinding pathways differ significantly among the inhibitors: KYP-2047 retains a preference for exiting through the tunnel in the $\beta$-propeller domain, even when mutations narrow the pathway, whereas HUP-46 strongly favors the inter-helical site, a behavior not observed for HUP-55. This distinction may arise from structural differences in the aminooxazole and aminothiazole moieties of the HUP ligands or from transient fluctuations in PREP’s side chains during the simulations.

A key observation was the phenomenon of ``pathway hopping,'' particularly evident in HUP-55, where the ligand diffusively explored multiple binding regions before selecting its exit. The umbrella sampling results also showed that transitioning between specific binding sites, such as from an alternative site to the active site for HUP-55, involves a relatively low free-energy difference of about 11 kJ/mol. This suggests that such movements could be a natural part of the physiological substrate processing of PREP.

The distinct binding behaviors of the HUP ligands compared to KYP-2047 raise questions about their potential role in modulating PPIs. The ability of the HUP ligands to interact dynamically with multiple binding sites might reflect an adaptation to the conformational flexibility of PREP, which could be integral to their PPI-targeting mechanism.

Our findings also highlight the complexity introduced by pathway hopping in systems with multiple unbinding mechanisms. When transition states for alternative unbinding pathways are energetically comparable, even minor changes to drug-protein interactions can switch the dominant pathway, complicating the design of inhibitors targeting specific kinetic properties. This highlights the significance of computational methods that can capture the entire range of binding and unbinding pathways, especially in drug discovery.

Although our free-energy estimates for KYP-2047 agree with experimental inhibition constants ($\ki$), the lack of corresponding $\ki$ values for the HUP ligands limits direct validation of their binding dynamics. More experimental validation is essential to understand better the binding dynamics of the HUP ligands and their implications for PREP functional dynamics. Such studies would provide critical insights into the mechanistic roles of these ligands and their potential applications in modulating PREP-mediated processes.

\begin{suppinfo}
Supporting Information is available free of charge at \url{https://pubs.acs.org/}.
\begin{itemize}
  \item Additional figures and tables that complement the main findings of the study.
  \item Detailed visualizations of ligand binding sites and analysis of residue interactions.
  \item Information about molecular dynamics simulations, including RMSF calculations and clustering analysis of ligand exit pathways.
\end{itemize}
\end{suppinfo}

\section*{Acknowledgments}
We thank Timo Myöhänen and Erik Wallén for the insightful discussions that shaped this work. J. R. acknowledges funding the Ministry of Science and Higher Education in Poland. We acknowledge Polish high-performance computing infrastructure PLGrid for awarding this project access to the LUMI supercomputer, owned by the EuroHPC Joint Undertaking, hosted by CSC (Finland) and the LUMI consortium through PLL/2023/04/016512. 
\bibliography{main}

\end{document}